\documentclass[10pt]{article}   
\usepackage{graphicx}

\begin{document}

\title{XORRO: Rapid Paired-End Read Overlapper}
\author{ Russell J Dickson$^{1}$
and
Gregory B Gloor\thanks{Corresponding Author: ggloor@uwo.ca}\\
 Department of Biochemistry, University of Western Ontario,\\ 
         London, ON, Canada
}
\date{}
\maketitle

\begin{abstract}
Background: Computational analysis of next-generation sequencing data is outpaced by data generation in many cases. In one such case, paired-end reads could be produced from the Illumina sequencing method faster than they could be overlapped by downstream analysis. The advantages in read length and accuracy provided by overlapping paired-end reads demonstrates the necessity for software to efficiently solve this problem.
              
Results: XORRO is an extremely efficient paired-end read overlapping program. XORRO can overlap millions of short paired-end reads in a few minutes. It uses 64-bit registers with a two bit alphabet to represent sequences and does comparisons using low-level logical operations like XOR, AND, bitshifting and popcount. 
        
Conclusions: As of the writing of this manuscript, XORRO provides the fastest solution to the paired-end read overlap problem.

XORRO is available for download at: 

sourceforge.net/projects/xorro-overlap/

\end{abstract}

\section*{Background}

Next-Generation sequencing platforms have increased the nucleotide-per-hour output such that the rate-limiting step of the analyses is computational rather than chemical. To keep up with the relentless pace set by high-throughput sequencing platforms such as Illumina, more efficient algorithms and implementations are needed. 
XORRO is an example of such software; it provides a solution to the paired-end read overlap problem and is practical for use on a 64-bit, single-CPU home computer. 

Overlapping paired-end reads provide increased practical read length and accuracy that allows for analysis of previous inaccessible genomic regions \cite{Caporaso:2011p685,Bartram:2011p686}. Recently, Gloor et al. \cite{Gloor:2010jy} demonstrated that paired-end Illumina reads could be used to profile the rRNA V6 region despite the fact that such reads were individually too short to cross the region, when profiling the vaginal microbiome \cite{Hummelen:2010p663}. This approach is becoming widely used \cite{Caporaso:2011p685,Bartram:2011p686,Gloor:2010jy,Hummelen:2010p663,Degnan:2012ce,Claesson:2010bh}. A key step in the analysis was the creation of paired-end reads that crossed the variable region of the rRNA V6 region to create a single overlapped read. 
Presently, sequencing technology providers do not provide software to perform the overlap step. Computationally, the actual overlap step is quite expensive because of the use of traditional Needleman-Wunsch-type string comparison methods.  XORRO provides a much faster algorithm for overlapping paired-end reads that are unlikely to have insertions or deletions, such as the reads produced by the Illumina sequencing platform.

 XORRO is the \emph{XOR}-based Read Overlapper, so-named because if its extensive use of the ``exclusive-or" (\emph{XOR}) logical operation for imperfect read matching. XORRO is able to overlap reads rapidly because it uses logical operations on bit strings representing nucleotide reads instead of directly overlapping the reads themselves.


\section*{Implementation}

\subsection*{Algorithm Overview}
XORRO is written in C and (for optional features including increased speed) x86\_64 Assembly Language. An overview of the XORRO algorithm is shown in Figure 1. The XORRO algorithm takes two ordered fastq files containing corresponding paired-end reads as input. The read pairs are converted from 8-bit characters to 2-bit binary representations. To find the correct longest between the sequences, the paired-end reads are compared sequentially starting from the longest possible overlap down to the shortest allowable overlap. When the longest match is found, a single overlapped sequence representing the conceptual read is returned. The overlapped read has a revised quality score equal to the sum of the two pairs used to make the overlapped region \cite{Cock:2010p553}. In the case of imperfect XORRO matching, the quality score is the difference between the two contradicting nucleotides. 

\begin{figure}[tbh]
\centerline{\includegraphics[width=\textwidth]{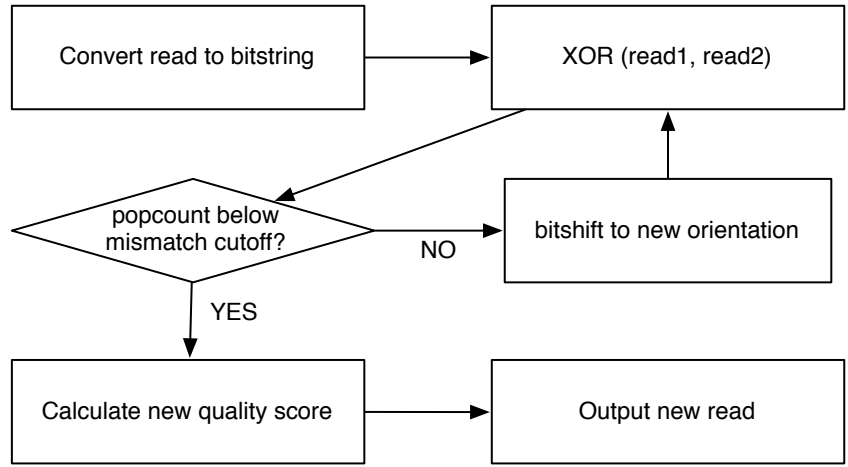}}
\caption{ Flowchart describing XORRO algorithm execution on a single pair of reads.  }
\label{flowchar}
\end{figure}

\subsection*{Bit Encoding}
Nucleotides are typically represented in a computer as an 8-bit character, as `A', `T', `G', and `C'; each comparison is done one-at-a-time. XORRO accelerates this process by representing each nucleotide as two bits of a 64-bit integer, which is the minimum number of bits required to represent characters in a 4-symbol alphabet. In this way, a two 64-bit integers representing 32 nucleotides can be compared in a single step. Because of the number of comparisons required, it is critical that the comparison step is optimized.

\subsection*{Imperfect Matching}
In XORRO, comparisons between reads are optimized to be done 32 positions at a time. This means that imperfect string matching (allowing some nucleotide mismatches) is by impossible doing direct comparisons; a single nucleotide mismatch and 32 mismatches are both `not equal' according to the processor with no further distinction. XORRO solves this problem by using the logical operation "exclusive or" (or XOR, its namesake). The XOR operation is a bitwise operation which returns 1 if there is a mismatch between the two input bits and a 0 if they are the same. Therefore, when XOR is applied to the two read bitstrings, the output is a string of bits representing the location of of every pairwise mismatch. 

Each overlapped pair must have less than the cutoff number of allowed mismatches. To calculate the number of mismatches, a popcount-like operation is applied to the XOR bitstring. \emph{popcount} is a processor instruction in the SSE4.2 instruction set which counts the number of `set' bits in a 64-bit value. The output of the \emph{popcount} instruction is the total number of nucleotide mismatches between the two nucleotides in the current overlap alignment. Thus verification of the current overlap can be done in a single instruction on a modern processor. For pre-SSE 4.2 CPUs, optimized popcount-like functionality is emulated in C.

\subsection*{Shifting}

If the comparison between the left and right paired-end reads does not yield a match, one read must be shifted relative to the other so subsequent comparisons can be made. XORRO uses bit shift and rotate operations to rapidly move the bit-encoded reads left or right in the processor registers. This extremely fast operation is written directly in assembly language and is performed without the need for additional counter variables to manage the relative location of the overlap. A non-assembly language build is also available.

\section*{Results and Discussion}

In our benchmarks, XORRO was able to overlap 3 gigabytes of paired-end reads in 4 minutes, while still allowing for imperfect matches. Before we created XORRO, the overlap step of the computational pipeline was the most time-consuming for our analyses. Figure 1 
shows that XORRO is nearly an order of magnitude faster than FLASh \cite{Magoc:2011dj}, and several orders of magnitude faster than another program, SHE-RA \cite{Rodrigue:2010p500}. Overlapping Illumina pair-end reads is a negligible part of the analysis pipeline if XORRO is used. 

\begin{figure}[tbh]
\centerline{\includegraphics[width=\textwidth]{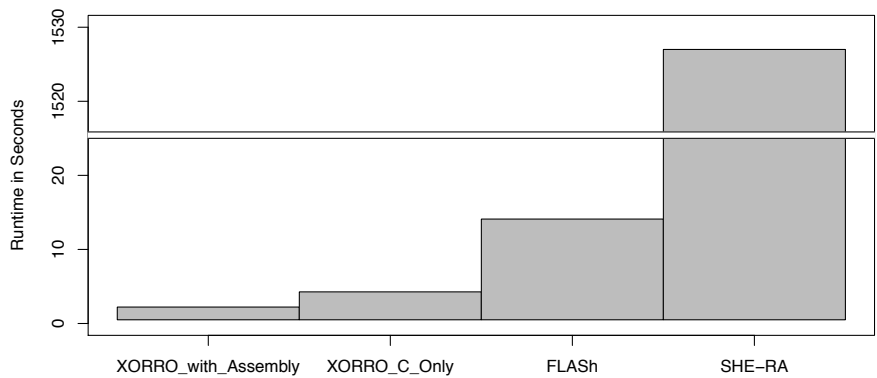}}
\caption{CPU runtime of published read overlapper programs. Overlapped 250 000 Illumina length 76 bp on a Core i7 Macbook Pro. `XORRO with Assembly' is XORRO with x64 assembly language calls for optimal speed. `XORRO C only' is XORRO without assembly language.}
\label{fig2}
\end{figure}

XORRO has been used for several recent publications to overlap paired-end Illumina reads \cite{Gloor:2010jy,Hummelen:2010p663} . A Perl wrapper has been included to automatically calculate parameters and provide a more user-friendly interface. Some longer paired-end read runs may extend beyond the range allowed by the 64-bit registers; in this situation multiple batch jobs may be required to find all pairs. The Perl wrapper can be used to easily implement such batch jobs; thus, typically users can allow the wrapper to automatically generate overlaps, while advanced users can customize their runs for greater efficiency. More detail can be found in the XORRO README instructions included in the package. 

The FASTQ format typically used to store read data is very similar to the FASTA format which is ubiquitous in bioinformatics  \cite{Cock:2010p553}. FASTQ contains a description line followed by the sequence data like FASTA; however, it also contains a nucleotide quality score which is represented by a single ASCII character (a letter, number, or symbol) corresponding with every nucleotide character. Each character is actually a numeric value that corresponds with a lookup table. Different standards exist for FASTQ files; because characters that are represented by low numbers are often reserved or invisible, most FASTQ formats do not start counting at zero and instead use an offset. It is important that the user is aware of which FASTQ score offset was used when producing their paired-end read files; this information should be provided as a parameter to XORRO if it is different than the default.

It is important to note that XORRO is not a general purpose string comparing program; it is optimized specifically to overlap short paired-end reads with a defined overlap length. One of the major differences between the XORRO algorithm and a general purpose string comparison algorithm is the omission of the ability to handle arbitrary insertions and deletions. Thus XORRO will not provide useful output when analyzing read data from technologies that are prone to indels. We considered a single mismatch to be acceptable during overlap; the output read description contains the number of mismatches.

XORRO is not based on the Needleman-Wunsch or \emph{k}-mer algorithms, as is common for nucleotide sequence overlap problems \cite{Rodrigue:2010p500}. Much of the speed advantage of XORRO can be attributed to bypassing the common overhead associated with such algorithms. Since XORRO was optimized with low-level programming operations, the algorithm can be performed with a minimal number of processor instructions. Thus XORRO is appropriate for use in high-throughput analysis pipelines using data generated from platforms with negligible indel frequencies.


\section*{Conclusion}
The increased length and accuracy of overlapped paired-end reads provide increased utility to next-generation sequencing methods. XORRO provides a simple solution to this computationally expensive problem.

\section*{Availability:}
XORRO is written in C (with some optional x86\_64 assembly language instructions). It requires a 64 bit x86-64 CPU. Multiple builds are available depending on CPU model. An optional Perl wrapper is included. XORRO is available for download at: 
sourceforge.net/projects/xorro-overlap/

\section*{Authors contributions}
GBG conceived the project. RJD created the implementation. RJD and GBG conducted experiments and wrote the manuscript.

\section*{Acknowledgements}
The authors would like to thank Jean M. Macklaim and Andrew D. Fernandes for their assistance in developing this manuscript.

 \bibliographystyle{bmc_article}  
  \bibliography{paperslibrary}     

\begin{figure}[!tpb]
\end{figure}

\end{document}